\documentclass[fleqn,usenatbib]{mnras}

\usepackage{newtxtext,newtxmath}

\usepackage[T1]{fontenc}
\usepackage{ae,aecompl}

\usepackage{amsmath}	
\usepackage{xcolor}
\usepackage{ulem}

\newcommand{\msun}{\hbox{M$_\odot$}}

\newcommand{\FeH}{\lbrack \rmn{Fe}/\rmn{H} \rbrack}
\newcommand{\kms}{\,{\rm km}\,{\rm s}^{-1}}
\newcommand{\kmsSq}{\,{\rm km}^2\,{\rm s}^{-2}}

\title[NSCs in the Galaxy]
{The Accreted Nuclear Clusters of the Milky Way}
\author[Pfeffer et al.]{
Joel Pfeffer,$^{1}$\thanks{E-mail: \href{j.l.pfeffer@ljmu.ac.uk}{j.l.pfeffer@ljmu.ac.uk}} 
Carmela Lardo,$^{2}$
Nate Bastian,$^{1}$ 
Sara Saracino$^{1}$ 
and \newauthor Sebastian Kamann$^{1}$
\\
$^{1}$Astrophysics Research Institute, Liverpool John Moores University, 146 Brownlow Hill, Liverpool L3 5RF, UK\\
$^{2}$Laboratoire d'astrophysique, Ecole Polytechnique F\'ed\'erale de Lausanne (EPFL), Observatoire de Sauverny, CH-1290 Versoix, Switzerland
}

\date{Accepted 2020 October 29. Received 2020 October 29; in original form 2020 May 20}

\pubyear{2020}

\begin{document}
\label{firstpage}
\pagerange{\pageref{firstpage}--\pageref{lastpage}}
\maketitle

\begin{abstract}
A number of the massive clusters in the halo, bulge and disc of the Galaxy are not genuine globular clusters (GCs), but instead are different beasts altogether.  They are the remnant nuclear star clusters (NSCs) of ancient galaxies since accreted by the Milky Way.  While some clusters are readily identifiable as NSCs, and can be readily traced back to their host galaxy (e.g., M54 and the Sagittarius Dwarf galaxy) others have proven more elusive.  Here we combine a number of independent constraints, focusing on their internal abundances and overall kinematics, to find NSCs accreted by the Galaxy and trace them to their accretion event.
We find that the true NSCs accreted by the Galaxy are: M54 from the Sagittarius Dwarf, $\omega$~Centari from \textit{Gaia}-Enceladus/Sausage, NGC~6273 from Kraken and (potentially) NGC~6934 from the Helmi Streams. These NSCs are prime candidates for searches of intermediate mass black holes (BHs) within star clusters, given the common occurrence of galaxies hosting both NSCs and central massive BHs.  No NSC appears to be associated with Sequoia or other minor accretion events.  Other claimed NSCs are shown not to be such.  We also discuss the peculiar case of Terzan~5, which may represent a unique case of a cluster-cluster merger.
\end{abstract}




\defcitealias{Massari_et_al_19}{MKH19}

\section{Introduction}
\label{sec:intro}

Nuclear star clusters (NSCs) are some of the densest and most massive clusters known, and occupy the central parts (i.e., nuclear region) of many galaxies.  For galaxies with stellar masses above $10^8$~\msun, the majority have readily identifiable nuclear clusters \citep[for a recent review, see][]{Neumayer20} and for galaxies with masses as low as $10^6$~\msun\ the occupation fraction of NSCs remains above $10$ per cent.  NSCs are unique amongst star clusters as they are the only type (including open clusters, globular clusters, and young massive clusters) that hosts clear evidence for multiple generations of star formation within them or extended star-formation histories \citep[SFHs; see the recent review by][]{Bastian_Lardo_18}.  While some NSCs display clear evidence of extended in-situ star-formation \citep[e.g.][]{Seth_et_al_06, Walcher_et_al_06, Feldmeier-Krause_et_al_15}, others may be formed by the merging of in-spiralling globular clusters \citep[GCs - e.g.,][]{cd93,lotz01} due to dynamical friction \citep{Tremaine_Ostriker_and_Spitzer_75}.

The result of these processes, either extended SFHs or the merging of GCs, is that NSCs are expected to host significant spreads in Fe-peak elements and/or extended SFHs.  Such spreads are readily observable in resolved colour-magnitude diagrams (CMDs) and/or spectroscopy, as well as with integrated spectroscopy \citep[e.g,][]{kacharov18}.  Heavy element spreads are rare amongst Galactic GCs, unlike the light elements (e.g., He, C, N, O, Na) abundance spreads, often referred to as ``multiple populations'' which appear to be universally present in old and massive stellar clusters \citep[e.g.,][]{gratton_et_al_12, martocchia_et_al_18}.

Due to the exquisite measurements provided by the \textit{Gaia} satellite, it has become possible to dynamically trace the origin of the majority of GCs within the Milky Way (MW - \citealt{Massari_et_al_19}, hereafter MKH19).  Using the integrals of motion, the authors were able to assign GCs to either an in-situ origin, or to one of the known galactic accretion events, i.e.,  \textit{Gaia}-Enceladus/Sausage (G-E/S), the Sagittarius dwarf galaxy, the progenitor of the Helmi streams, and the Sequoia galaxy.  Additionally, they found a collection of GCs, which they term the "Low-energy" group, without (at the time) any known accretion event.  This group is likely to be associated with the Kraken merger event, an early and relatively massive accretion event that has been postulated based on the kinematics and age-metallicity distribution of GCs \citep{Kruijssen_et_al_19a,Kruijssen_et_al_19b, Kruijssen_et_al_20,Pfeffer_et_al_20,Forbes_20}.

The inferred stellar masses of these accreted dwarf galaxies ($\sim10^8$~\msun) imply that many of them would be expected to host an NSC, based on the occupation fraction distribution observed today \citep{sanchez_janssen19}.  The goal of the present work is to use the updated sets of constraints available to uncover the NSCs brought in as part of these accretion events and assign them to their host galaxy.

There has been much previous work done to uncover the accreted NSCs within the Milky Way GC population \cite[e.g., see the review by][]{da_costa_16}.  The most clear cut case in the Galaxy is that of the GC M54, which hosts an extended star-formation history and large Fe-spread \citep[e.g.,][]{alfaro_et_al_19} and is physically associated with the central regions of the Sagittarius dwarf galaxy \citep{Ibata_et_al_95}.  However, as will be explored later in the paper, even in this clear case there has been some confusion about the actual abundance spread (see \S~\ref{sec:m54}).  The best studied example of a GC hosting a large Fe-spread that is likely an accreted NSC is $\omega$~Centauri \citep[e.g.,][]{Lee_et_al_99, Hilker_and_Richtler_00, bekki_freeman_03}, although which accretion event it is associated with is still debated.

While finding Galactic GCs with Fe-spreads is a promising and powerful way to identify accreted NSCs within the Galaxy, the picture has been somewhat complicated by the growing number of Galactic GCs with alleged Fe-spreads \citep[see][]{da_costa_16}.  If all of these GCs would be accreted NSCs, then we would be missing a large fraction of galactic accretion events (of massive satellites), which is unlikely given the \textit{Gaia} results to date \citep[e.g.,][]{helmi_20}.  However, follow-up studies of a number of clusters with claimed Fe spreads suggest that some may not be true, as the abundances measured from Fe~{\sc I} and Fe~{\sc II} give conflicting results \citep[e.g.,][]{mucciarelli_et_al_15}.  In the present work we will critically assess the claims of Fe spreads present in Galactic GCs, and for the likely NSC candidates we will attempt to associate each one with their host (accreted) satellite galaxy.

This paper is organised as follows: in \S~\ref{sec:constraints_iron} we first discuss constraints obtained through abundance spreads.  The kinematical constraints are explored in \S~\ref{sec:constraints_kinematics}.  Finally, in \S~\ref{sec:analysis} we combine all of the constraints in order to identify the accreted NSCs within the Milky Way, as well as which accreted satellite they were brought in with.

\section{Constraints from Iron Spreads}
\label{sec:constraints_iron}

\subsection{Background on Spectroscopic Methods}

There are a number of ways that spreads in the [Fe/H] content in stars within a GC can be inferred. There have been several studies that used relatively low-resolution spectra of a number of stars within a GC and looked for variations in the CaT lines, as a proxy for [Fe/H] spreads \citep[e.g.][]{Husser_et_al_2020}. This can be an efficient way to find such spreads, but a number of caveats exist so some care must be taken.  The most direct way to infer spreads in [Fe/H] is through the analysis of high-resolution spectra (typically done for red giant branch - RGB - stars) to measure [Fe/H] directly.  However, even this direct method has some caveats, as in some cases (which will be discussed below) the measurements for Fe~{\sc i} and Fe~{\sc ii} disagree.

\subsection{Background on Photometric Methods}

Recently, due to the extensive survey of Galactic GCs with the Hubble Space Telescope \citep{piotto_et_al_15}, exquisite photometry for thousands of GC members exists from the UV to the optical.  \citet{milone17} have combined the photometry into pseudo-colours to create the so-called ``chromosome maps".  These maps are extremely efficient in
separating out the light element abundance patterns, a.k.a. ``multiple populations", with the P1 (normal, or field-like) stars offset from the P2 populations (those with enhanced N, Na, and Al and depleted O and C).

However, when constructing the chromosome maps, \citet{milone17} found that while most clusters were similar, containing versions of the P1 and P2 populations (which they refer to as Type I clusters), a number of clusters (referred to as Type II) showed additional populations.  Most of these Type II clusters showed an additional population of stars above and to the right of the nominal P2 stars in the chromosome maps.  We will refer to these stars as the P2-anomalous population.  Additionally, the true Fe-spread clusters (see below) also contained stars that were offset from the main P1 population (which we will refer to as the P1-anomalous stars).

As every GC studied in the necessary detail to date displays the P1 and P2 populations, a merger of two clusters (one of the pathways to form an NSC, likely to be the dominant process at dwarf galaxy masses - \citealt{Neumayer20}) should show a spread in [Fe/H] as well as the two sub-populations P1 and P2.  Likewise, if a GC is able to accrete material and form a second generation of stars within it, it would be expected to form a P1 population (and perhaps a P2 populations) that is offset in [Fe/H] from the existing P1 and P2 populations.  These stars would sit to the right and slightly down in the chromosome map if they are enhanced in [Fe/H] due to their cooler temperatures \citep[e.g.,][]{marino_et_al19}.  Hence, we will use the available chromosome maps to look for both the anomalous P1 \& P2 in determining if [Fe/H] spreads due to second generations of star formation or GC mergers were at play in the formation of the present day GC.

The origin of the anomalous P2 populations within the Type~II clusters without a corresponding anomalous P1 population is still unknown.  The P2 anomalous stars are generally enhanced in the C+N+O sum (where this sum is consistent with being constant in most other clusters/populations) and enriched in s-process elements \citep[see e.g.,][]{Bastian_Lardo_18}.  In some cases, these P2 anomalous populations have led to the suggestion that spreads in [Fe/H] are present within the cluster, which will be discussed in detail in \S~\ref{sec:chemistry_individual}.

Unless otherwise noted, all of the chromosome maps discussed in this paper are from \citet{milone17}.

\subsection{Individual Clusters}
\label{sec:chemistry_individual}

\subsubsection{NGC~5139 ($\omega$~Cen)}

The best studied GC with a clear [Fe/H] spread is $\omega$~Cen, which hosts a $\sim1.2$~dex spread in iron \citep{norris95,norris_omega,suntzeff_omega,johnson10}.  The dominant population has $\FeH\sim-1.8$ with subsequent minor peaks at $-1.5$, $-1.1$ and $-0.8$.
Spectroscopic evidence for a centrally concentrated population of very metal-poor stars (between $\FeH = -2.30$ to $-2.52$) was recently reported by \citet{johnson_et_al_2020}. However, no such population is evident in the study of \citet{Husser_et_al_2020}, despite the larger sample size of their study. If confirmed, this population would be near the observed floor in GC metallicity in the Milky Way and other Local Group galaxies \citep{Beasley_et_al_19, Kruijssen_19}.

The chromosome map of NGC~5139 is one of the most complicated observed to date, clearly reflecting the presence of significant Fe spreads \citep{milone17}.

\subsubsection{NGC~6715 (M54)}
\label{sec:m54}

\citet{carretta_et_al_10} have presented [Fe/H] measurements for a large sample of stars within M54 and report an Fe-spread with a dispersion of $\sigma=0.19$~dex.  However, closer scrutiny reveals that this spread only refers to the oldest populations within the cluster.  When the young and intermediate age populations are also included, an Fe-spread of nearly 1.5~dex becomes apparent \citep{carretta_et_al_10,carretta_m54,alfaro_et_al_19}.  The associated age spread of the full stellar population of this cluster covers $\sim10$~Gyr, from $\sim2$~Gyr ago to $\sim12$~Gyr ago \citep[e.g.,][]{bellazzini_et_al_08,alfaro_et_al_19}.

When studying the internal metallicity spread of M54, one needs to account for the contribution of field stars from the Sagittarius dwarf, given the location of the cluster inside the stream of the disrupting galaxy. Indeed, \citet{alfaro_et_al_19} found that the intermediate-age population is the least centrally concentrated of the three, and hence could be composed of Sagittarius field stars. On the other hand, the young metal-rich populations appears to be the most centrally concentrated, and hence clearly associated with M54. This is also supported by the kinematics of the populations \citep[see][]{alfaro_et_al_2020}.

The chromosome map of NGC~6715 is correspondingly complicated, due to the spread in [Fe/H] as well as in age.  Each of the two old stellar populations within the cluster display the characteristic properties of multiple populations \citep{carretta_m54}.  While the intermediate age ($3-9$~Gyr) and young components ($\sim2$~Gyr) do not display strong evidence for MPs \citep{sills19}, this may be due to the observed correlation between a populations age and the strength of their MPs \citep[e.g.,][]{martocchia_et_al_18,saracino_et_al_20} or association with the field star population of Sagittarius \citep{alfaro_et_al_19}.

\subsubsection{NGC~6273 (M19)}

\citet{johnson_m19} showed that NGC~6273 hosts three distinct stellar populations with an extended metallicity distribution. The metal-poor group has $\langle \FeH\rangle =-1.75$ ($\sigma =0.04$), whereas the metal-rich component has $\langle \FeH \rangle  = -1.51$ ($\sigma = 0.08$) and enhanced s-process content. They also detected a possible anomalous group with $\langle \FeH\rangle =-1.30$ (1 star) and noticeably lower [X/Fe] ratios for nearly all elements. The two dominant populations are nearly equivalent in number, whereas the anomalous group constitutes only 6 per cent of the whole spectroscopic sample. The authors also note that other clusters (i.e. M2 and NGC~5286) host a minority population of anomalous stars with peculiar abundances that may indicate that these stars were originally part of a different system or accreted from a larger progenitor host.

These results were later confirmed by \citet{yong_et_al_16}, who analysed a large sample of CaT spectroscopy of this cluster and found a range of [Fe/H] values spanning $\sim1$~dex ($\sigma$= 0.17).  This was followed up by \citet{johnson_et_al_17} who used a large sample of high and medium resolution spectra of NGC~6273 members, and found a dispersion in [Fe/H] of $\sim0.2$~dex, with a full range of values from $\FeH \sim -2$\ to $-1$~dex\footnote{We note that while \citet{johnson_m19,johnson_et_al_17} used the `spectroscopic' approach to determine the Fe spread within M~19, which can lead to an artificial broadening of the intrinsic Fe spread \citep[e.g.,][]{Lardo_et_al_16,Mucciarelli_16}, the large spread found in M~19 ($\sim1$~dex) argues that a large and true Fe spread exists within this cluster. Future work using the ``photometric'' method on this cluster will be able to confirm this.}.  Additionally, they find some evidence for three discrete sub-populations in metallicity space with each sub-populations exhibiting the classic signatures of MPs. They also identified a population of at least five peculiar ``low-$\alpha$'' stars that have $\mathrm{[\alpha/Fe]} \sim 0.0$ and low [Na/Fe] and [Al/Fe] abundances.

\subsubsection{Terzan~5}

\citet{ferraro_et_al_09} found evidence, based on a relatively small sample of stars, for a significant age and metallicity spread within Terzan~5.  \citet{origlia_et_al_11} provided an analysis of a much larger sample and confirmed that the cluster was made up of two major components, one with $\FeH \sim -0.25$ and another with $\FeH \sim +0.27$.  The lower metallicity group is also significantly enhanced in [$\alpha$/Fe] in line with expectations of galactic chemical enrichment if there is a significant age difference between the populations. 
More recently, \citet{massari_et_al_14} confirmed the previously identified peaks in [Fe/H] and also identified a metal-poor component at $\FeH \sim -0.8$~dex, which accounts for $\sim6$ per cent of the stars in the cluster. 

\citet{ferraro_et_al_16} found evidence for two discrete main sequence turn-offs within the cluster, suggesting a $\sim8$~Gyr delay between the formation of the dominant sub-solar component and the super-solar metallicity component.

Both its large metallicity spread and the absence of a measurable Al-O anti-correlation in each metallicity group have been used as arguments for Terzan~5 not being a ``true'' globular cluster \citep{origlia_et_al_11}. However, variations in [Al/Fe] and [O/Fe] are expected to be extremely small in metal-rich globular clusters \citep{pancino17,nataf19}. Also, the five most metal-poor stars presented in \citet{nataf19} with metallicities between [Fe/H] =--0.66 and --0.46 show the usual CNO abundance variations, with sodium and potassium variations possibly detected as well. Thus, the light-element abundance variations among Terzan 5 stars are consistent with those seen in normal globular clusters at the same metallicity \citep{nataf19}.

While the evidence for significant age and [Fe/H] spreads is quite clear in Terzan~5, it remains a distinctive case which we will discuss in more detail in \S~\ref{sec:analysis_terzan5}.

\subsubsection{NGC~6934}

\citet{marino_et_al_18} studied 4 stars with high resolution spectra located on the split RGB of NGC~6934. They found intrinsic Fe variations with a difference in iron of the order of $\sim$0.2 dex. Importantly, the authors did not find evidence of s-process spreads. Given the lack of any star-to-star variations in  s-process elements of NGC~6934 stars associated with the inferred changes in metallicity (as in the case for i.e. M~22 and NGC~1851; see below), we conclude that the measured [Fe/H] spread is likely real.

Looking at the chromosome map of this cluster, it does show a small population of stars with enhanced [Fe/H] that contains both the nominal P1 and P2 sub-populations \citep[$\sim$6.7 per cent,][]{milone17}.  Hence, we conclude that NGC~6934 likely hosts an [Fe/H] spread and may be an NSC.

\subsubsection{NGC~2419}
\label{sec:n2419}

NGC~2419 is among the most luminous ($\mathrm{M}_{V} = -9.5$; \citealp{bellazzini_2419}) and massive ($M \simeq 10^6$~\msun;  \citealp{ibata_2419}) GCs in the halo of the Milky Way.  It is located at a galactocentric distance of d = 87.5 kpc \citep{dicriscienzo_11}. 
Its half-light radius is significantly larger than that of other GCs with similar luminosity, more akin to the nuclei of dwarf galaxies than to classical GCs \citep{vdb04}.
This has led several authors to suggest that NGC~2419 could be the stripped core of a former dwarf galaxy\citep[e.g.;][]{mackey_2005}.

\citet{Cohen_2010} measured [Ca/Fe] abundances for 43 bright giants in NGC~2419 from moderate resolution spectra around the calcium triplet. They found a significant spread in their inferred [Ca/H] values, with a prominent peak at $\mathrm{[Ca/H]} \simeq -1.95$ and a dispersion of $\sim0.2$~dex, which was larger than the measurement errors. \citet{Cohen_2011} followed up this result with high resolution spectra of seven luminous RGB stars of NGC~ 2419, finding that NGC~2419 stars do not display any spread in [Fe/H] or [Ca/Fe] in excess to what is expected from the observational uncertainties.
Additionally, they observed a star with a very low [Mg/Fe] ratio, but being normal in all other element ratios except for a high [K/Fe]. This chemical anomaly was never reported before for GC stars \citep{Cohen_2011}. 

The lack of any intrinsic spread in [Fe/H], [Ca/H], and [Ti/H] was further confirmed by \citet{mucciarelli_2419}, who also demonstrated that NGC~2419 indeed exhibits a large dispersion in Mg abundances ($>1$~dex).   According to \citet{mucciarelli_2419}, such a large spread in Mg abundances is the cause of the dispersion in [Ca/H] observed by \citet{Cohen_2010}, as the observed severe Mg depletion leads to an increase of the equivalent widths of the Ca triplet lines at a constant Ca abundance. Indeed, the strength of the CaT lines 
is sensitive not only to the abundances of Ca and Fe, but also to the abundance of those elements that affect the H$^{-}$ continuum opacity (e.g. $\alpha$-elements), through their contribution to the free electronic density.

A large Mg depletion might in principle cause an apparent range in CaT line strengths that mimics a small abundance spread also in other clusters. However, significant star-to-star variations in Mg abundance have been found to date only in a few cases, mostly metal-poor and massive GCs \citep[see][for a discussion]{Carretta2014}. Thus, the case of NGC~2419 is clearly unique among stellar clusters in the Milky Way.

\subsubsection{NGC~5824}

\citet{da_costa_et_al_14} carried out a low-resolution (but large sample) study of this cluster, using the CaT as a proxy for [Fe/H].  They suggest that an iron spread likely exists within the cluster.  
\citet{roederer_et_al_2016} performed the first high-resolution study of a large sample of RGB stars in NGC~5824. In particular, they presented a detailed abundance analysis of 17 elements for 26 stars.  The authors measured 
$\langle [\mathrm{Fe}/{\rm{H}}]\rangle =-1.94 \pm 0.02$ (statistical) $\pm$ 0.10 (systematic). From their analysis, they were able to exclude the presence of an intrinsic metallicity spread at the 0.08 dex level.
Similarly, \citet{mucciarelli_18} followed up this result with high resolution spectroscopy and did not find a substantial spread in [Fe/H]\footnote{In particular, \citet{mucciarelli_18} measured a mean metallicity of $\FeH = -2.12 \pm 0.01$ with an intrinsic scatter of $0.00 \pm 0.02$ from the analysis of the 66 RGB stars in common with \citet{da_costa_et_al_14}.}.
NGC~5824 also displays a large range of [Mg/Fe] abundance, observed only in a few metal-poor and/or massive clusters \citep{mucciarelli_18}. The [Fe/H] abundances (as derived from \ion{Ca}{II} lines) are also mildly anti-correlated with [Mg/Fe], in the sense that NGC~5824 stars with low [Mg/Fe] have systematically higher [Fe/H] abundances. This led \citet{mucciarelli_18} to conclude that such an unusual Mg depletion (down to $\mathrm{[Mg/Fe]} \sim -0.5$) gave rise to a significant increase of the EWs of the 
\ion{Ca}{II} lines and that can be erroneously interpreted as a high Fe abundance.
Thus, the metallicities inferred from \ion{Ca}{II} lines could be over-estimated in Mg-poor stars, as in the case of NGC~2419.

\subsubsection{NGC~6656 (M22)}

This cluster has been the subject of some debate about whether it hosts [Fe/H]-spreads within it. 
\citet{hesser_et_al_1977} were the first to note the similarities of M22 and $\omega$~Cen. \citet{norris_et_al_1983} showed that the CH and CN variations in M22 were correlated with \ion{Ca}{ii} H and K line variations, indicating that both C and N are overabundant in a high fraction of the stars in M22, which also appear to be enriched in Ca (i.e. more metal-rich). Numerous studies of this cluster over the last decades have yielded conflicting results: depending upon the sample and the adopted analysis techniques, some authors measure no significant variations whereas others find significant iron variations up to $\sim$0.5~dex \citep[e.g.,][]
{cohen_1981,gratton_1982,lehnert_et_al_1991,brown_1992}. More recently, \citet{dacosta_et_al_2009} analysed intermediate resolution spectra around the \ion{Ca}{ii} triplet to trace Fe variations and found an iron abundance distribution that is substantially broader than that expected from the observed errors alone, with a peak at $\FeH \sim -1.9$ and a broad tail to higher abundances. The authors also note that the abundance distribution among M~22 stars bears a qualitative similarity to that for $\omega$~Cen, although the ranges of the chemical variations in M22 are considerably smaller.

Based on a sample of high-resolution spectra, \citet{marino_et_al_09} report the presence of two sub-populations, with a difference in their [Fe/H] content of $\sim0.14$~dex.  They also found a significant spread in s-process elements, with stars with higher [Fe/H] values also having large s-process abundances.

\citet{mucciarelli_et_al_15} argued that the intrinsic iron spread measured from high-resolution spectra by \citet{marino_et_al_09} was due to differences in the measured values of Fe~{\sc i} and Fe~{\sc ii} (which should ideally give the same results; see also \citealp{ivans_et_al_2004}). In particular, \citet{mucciarelli_et_al_15}  re-analysed the sample of 17 RGB stars discussed in \citet{marino_et_al_09}.  In contrast to \citeauthor{marino_et_al_09}, who derive atmospheric parameters following a standard fully spectroscopic approach, \citet{mucciarelli_et_al_15} used two different methods to constrain effective temperatures and surface gravities. When atmospheric parameters are derived spectroscopically, they measure a bimodal metallicity distribution, that well resembles that by \citet{marino_et_al_09}. However, the metallicity distribution from \ion{Fe}{II} lines strongly differs from the distribution obtained from \ion{Fe}{I} features when photometric gravities are adopted. The \ion{Fe}{I} distribution still mimics the [Fe/H] distribution obtained using spectroscopic parameters, whereas the \ion{Fe}{II} shows the presence of a single stellar population which is internally homogeneous in iron.
The authors suggest that the difference may be caused by non-local thermodynamical equilibrium (NLTE) effects or over-ionisation mechanisms, and such differences have now been found in other clusters.  Interestingly, in all such cases, the GCs show spreads in their s-process elements, which is atypical for GCs. Indeed, the analysis presented in \citet{mucciarelli_et_al_15} confirms the presence of the two s-process element groups found by \citet{marino_et_al_09}. The significant range in Ca II triplet line strengths seen among the red giants in M~22 remains to be explained. Indeed, because of the relatively modest Mg variations and the higher overall metallicity of the cluster, it seems unlikely that star-to-star Mg variations are driving the Ca triplet variations observed by \citet{dacosta_et_al_2009}.

Finally, in contrast to the results of \citet{alves_et_al_2012}, who argued for a $\sim$0.4 dex intrinsic iron spread in this cluster based on high-resolution infrared spectra, \citet{Meszaros_20} did not find any compelling evidence for significant Fe variations in M22, from the spectroscopic analysis of 80 RGB stars from the SDSS-IV APOGEE-2 survey.

The chromosome map clearly shows an anomalous population of stars above the standard P2 population.  However, no such stars are seen corresponding to an Fe-enriched P1 population.  This is further evidence against actual [Fe/H] spreads within the cluster. 

We conclude that M~22 is not likely to host significant spreads in [Fe/H], hence it is not a candidate NSC.

\subsubsection{NGC~1851}

NGC~1851 is a relatively massive GC characterised by a double sub-giant branch (SGB) in its colour-magnitude diagram \citep{milone_et_al_2008}. Its shows a range in C+N+O abundance among RGB stars \citep[e.g.,][]{yong_et_al_2015} and star-to-star variations in
s-process elements \citep{yong_et_al_2008,carretta_et_al_11,lardo_et_al_2012, gratton_et_al_2012}.
The cluster is surrounded by a symmetric, diffuse stellar halo which extends more than 250~pc in radius, with no evidence of tidal tails (\citealp{olszewski_et_al_2009}; but see \citealp{sollima_et_al_2012,carballo_et_al_2018}).
The presence of this stellar system surrounding NGC~1851 has lead some to speculate that the cluster might have been originally formed in a dwarf galaxy which is now tidally disrupted \citep{marino_et_al_2014,simpson_et_al_2017,kuzma_et_al_2018}.

\citet{carretta_et_al_11} measured the [Fe/H] abundance, along with a number of other elements, in 124 giant stars within this cluster.  They report a spread in [Fe/H] of $\sigma=0.07$~dex which is larger than their nominal uncertainties in the measurements of individual stars ($\sim0.03$~dex).  The authors find a larger abundance spread if they analyse Fe and Ba together, as the nominally Fe-rich population are also rich in Ba.  One potential problem with this analysis is that Ba is an s-process element, and clusters that show s-process element abundance variations have been associated with spurious claims of Fe spreads \citep[see also][]{Husser_et_al_2020}. \citet{villanova_et_al_2010} presented a chemical abundance analysis of a sample of 15 RGB stars in this cluster. They found that the Ba distribution is bimodal, whereas the iron abundance for the two Ba groups is the same within the errors.
Given the small (and still disputed) [Fe/H] spread and the associated s-process variations, we do not consider NGC~1851 to be a strong candidate accreted NSC.

\subsubsection{NGC~7089 (M2)}
\label{sec:M2}

A case that is similar to M22 and NGC~1851 is that of M2. 
\citet{yong_et_al_14} analysed CaT and high-resolution spectroscopy of a large sample of RGB stars within M2 and inferred a significant spread in [Fe/H] with three peaks;  a large peak at $\FeH = -1.7$ and two smaller metallicity components at $\FeH = -1.5$ and $-1.0$.

\citet{Lardo_et_al_16} followed up these results and found a similar behaviour as was found in M22 for the two main peaks \citep{mucciarelli_et_al_15}.  Specifically, that the two peaks at $\FeH = -1.7$ and $-1.5$ merged into a single peak when using photometric gravities and treating \ion{Fe}{I} and \ion{Fe}{II} separately.  However, the small sub-population at $\FeH = -1.0$ remained. This population makes up $\sim1$ per cent of the stellar population in the cluster.
The presence of the sparse, metal-rich component, which shows neither evidence for sub-populations nor an internal spread in light-elements, is confirmed by the study of \citet{milone_et_al_2015_m2}.

Due to the extremely low fraction of [Fe/H] enhanced stars, we do not consider M2 to be a strong candidate to be a NSC, and instead suggest that these stars may constitute a rare accretion event from one GC to another \citep[e.g.,][]{khoperskov_et_al_18} or the accretion of a small cluster on M2.

However, the origin of this cluster (and its large Fe spread within a small minority of cluster members) is still uncertain as the likelihood of each investigated formation channel is low.

\subsubsection{NGC~362}

NGC~362 was observed at high-resolution by \citet{carretta_13}.
They collected spectra for 92 RGB stars in this cluster and found no evidence for an internal metallicity dispersion \citep[see also][]{Meszaros_20}. Moreover, \citet{carretta_13} discovered the presence of an additional, poorly populated (e.g. accounting for only $\sim$6 per cent of the total cluster population), red RGB sequence which appears to be enriched in s-process elements; similarly to that observed in M~22, NGC~1851, and in the bulk of giant stars in M~2. 
More recently, \citet{Husser_et_al_2020} measured a small metallicity variation of $\sim$0.12 dex among NGC~362 stars. This result is based on low resolution spectra centered around the Ca triplet feature. Even if the observed dispersion in metallicity is statistically significant, the latter result is based on the analysis of 22 stars in the s-process rich population (i.e. the population of anomalous P2 stars present above the nominal P2 population in the chromosome map).  For comparison, the authors observed 797 stars  along the main RGB body. 

Overall, the observational evidence would support the notion that NGC~362 is not characterised by an intrinsic iron spread and that the slightly higher [Fe/H] abundances observed in the s-process rich group of stars is likely to be introduced artificially by how atmospheric parameters and metallicities are derived in the spectroscopic analysis \citep[e.g.][]{mucciarelli_et_al_15,Lardo_et_al_16}.

\subsubsection{NGC~5286}

\citet{marino_et_al_n5286} report an [Fe/H] spread in this cluster, with two peaks separated by $0.2$~dex, along with significant variations in s-process elements.  This is very similar to what was found for M2 and the two main populations of M22.
The analysed sample includes stars observed at different resolution with UVES and GIRAFFE. Unfortunately, Fe abundances from neutral and ionised Fe are only available for the UVES spectra, which offers both higher resolution and larger spectral coverage.
Similarly to the case of M~22, when spectroscopic gravities are used for the analysis of UVES spectra, the distributions of [\ion{Fe}{I}/H] and [\ion{Fe}{II}/H] are very similar and broad, pointing to an intrinsic iron scatter. On the other hand, the metallicity distribution derived from \ion{Fe}{II} lines and photometric gravities is narrow and points out a lack of iron spread \citep{Mucciarelli_16}.

Like was the case for M2, the chromosome map of NGC~5286 shows the anomalous population above and to the right of the nominal P2 sub-population, typical of the "Type II" clusters \citep{milone17}.  However, in this cluster we do not see the corresponding anomalous P1 stars that would be expected if the [Fe/H] spread was real.  

Hence, we conclude that NGC~5286 is unlikely to host a significant [Fe/H] spread within it.

\subsubsection{NGC~6864 (M75)}

\citet{kacharov_et_al_13} used high resolution spectroscopy of 16 giant stars within NGC~6864 to derive detailed abundances.  They report a small [Fe/H] spread ($\sigma = 0.07$~dex).  Like the clusters discussed above with small [Fe/H] spreads (NGC~1851; M2) this cluster appears to also have spreads in s-process elements (e.g., Ba).  Like with M2, M22 and NGC~5286, we do not consider NGC~6864 to be a strong NSC candidate.

\subsubsection{NGC~3201}

\citet{simmerer_et_al_13} report a spread of $\FeH = 0.4$~dex from minimum to maximum.  Using the measurements presented in their paper we find a dispersion in [Fe/H] of $0.10$~dex which is only marginally larger than their average error on an individual measurement ($0.09$~dex).  
\citet{mucciarelli_et_al_3201} re-analysed their spectra to show that the metal-poor component claimed by \citet{simmerer_et_al_13} is composed by asymptotic giant branch stars that could be affected by NLTE-effects driven by iron overionization. Such NLTE effects have an impact on the iron abundances measured from Fe~{\sc I} lines (by $0.1-0.2$~dex), but leave the abundances from Fe II lines unchanged. Thus, \citet{mucciarelli_et_al_3201} conclude that the observed iron spread is not intrinsic but rather due to the inclusion of  AGB stars in the sample \citep[see also][]{ivans_et_al_2001,lapenna_et_al_2014,lapenna_et_al_2016}.
Hence, we conclude that this GC does not host large [Fe/H] spreads.

\subsubsection{NGC~6229}

\citet{johnson_et_al_17} report a small spread in [Fe/H] ($\sigma=0.06$~dex) in the massive outer halo cluster NGC~6229.  The authors note that such a spread is only marginally significant with respect to the uncertainties.  They compare the [Fe/H] distribution to NGC~1851 and find them to be similar.  However, as discussed above, NGC~1851 likely does not host a significant spread in [Fe/H].  The authors find that, like NGC~1851, NGC~6229 also shows s-process variations.  Hence, it is likely to be a Type II cluster, although a chromosome map does not currently exist for this cluster.

We conclude that, like NGC~1851, NGC~6229 is unlikely to be an accreted NSC.

\subsubsection{NGC~6388}

We found an additional cluster in the \citet{milone17} catalogue, that based on its chromosome map, may be an additional candidate NSC.  This is NGC~6388, a Type~II cluster with a number of stars that may represent an anomalous P1 population.  However, we note that this bulge cluster displays significant differential extinction.  The apparent anomalous P1 population overlaps with the nominal P1 population and is extended along the reddening vector.  

\citet{carretta_et_al_2007} presented a detailed chemical analysis of this peculiar bulge cluster based on the analysis of high resolution spectra of seven RGB stars. They found an average value  $\mathrm{[Fe/H]}=-0.44\pm0.01\pm0.03$ dex with no evidence of intrinsic spread in metallicity. The absence of any star-to-star Fe variations was also confirmed in subsequent studies from the same authors based on larger sample of giant stars observed at high-resolution \citep[see][and references therein]{carretta_et_al_6388}. Through low-resolution spectra, \citet{Husser_et_al_2020} recently analyzed a sample of RGB stars, claiming for the presence of a metallicity spread of $\sim$ 0.22 dex in the cluster. This finding is based on stars lying on top of the C+N+O and likely s-process rich population in the chromosome map, and, as stated by the authors, it may be affected by problems with the underlying photometry\footnote{Being located in the Bulge of the Galaxy, the photometry for this cluster is very problematic. This directly affects the extraction process of the spectra from the MUSE data cubes and therefore the quality of the spectra. As a consequence, large uncertainties are associated to the measured metallicities.}.

Hence, until further confirmation, we conclude that this is likely a Type II cluster without an Fe spread.

\section{Kinematical Constraints}
\label{sec:constraints_kinematics}

Based on the above literature review, we find that there are six clear cases of large [Fe/H] spreads within the MW GC population.  They are \citep[in order of mass,][]{Baumgardt_and_Hilker_18}: $\omega$~Cen, M54, NGC~6273, Terzan~5, NGC~7089 and NGC~6934.  In this section we will use kinematical information to associate these NSC candidates with MW accretion events.

\subsection{Individual Clusters}

\subsubsection{NGC~5139 ($\omega$~Cen)}

$\omega$~Cen has been recently suggested to be the former nuclear star cluster of either G-E/S or Sequoia. \citet{Myeong_et_al_19} suggest it is associated with the Sequoia accretion event based on the actions, inclination and eccentricity of its orbit, while \citetalias{Massari_et_al_19} suggest it is associated with G-E/S based on the binding energy of its orbit within the MW. Similarly, \citet{Forbes_20} assigned $\omega$~Cen and NGC 1851 as the NSCs of Sequoia and G-E/S, respectively.

Though there is significant overlap with the G-E/S debris, Sequoia field stars have typical energies $E \approx -1$ to $-1.3 \times 10^5 \kmsSq$ \citep{Myeong_et_al_19} and a nuclear star cluster of Sequoia would be expected to initially have a similar orbital energy, as galaxy accretion events deposit their GCs over a limited range in energies \citep{Pfeffer_et_al_20}. Thus, dynamical friction would need to act to reduce $\omega$~Cen to its present orbit with $E \approx -1.85 \times 10^5 \kmsSq$, or its apocentre would need to be reduced from an initial $\sim 30$~kpc to its current value of 7~kpc.
In contrast, dynamical friction could act on the host galaxy to deliver $\omega$~Cen to near its present apocentre \citep{bekki_freeman_03}, however the stellar debris from the merged galaxy would then be found at similar (small) apocetres.

Following \citet[their Appendix B]{Lacey_and_Cole_93}, we calculate the dynamical friction timescale for the potential of an isothermal sphere, assuming a circular velocity of $V_c \approx 230 \kms$ for the Milky Way\footnote{Any reasonable value for the circular velocity does not affect these results since the dynamical friction timescale approximately scales with $V_c$. Thus the results are similar when adopting smaller circular velocities for the Milky Way at $z>0$.} \citep{Bland-Hawthorn_and_Gerhard_16} and adopting an orbit circularity of $0.5$, similar to $\omega$~Cen.
For its present-day mass of $3.5 \times 10^6$~\msun\ \citep{Baumgardt_and_Hilker_18}, it would take $>300$~Gyr to reduce the apocentre of $\omega$~Cen from 30 to 7~kpc. For this to occur within a Hubble time, $\omega$~Cen would need a mass $\approx 50$ times its present mass (i.e. a mass similar to that suggested for its host galaxy) without suffering mass loss through tidal effects.
Similarly, starting at 20~kpc rather than 30~kpc still implies a timescale $\approx 130$~Gyr for the present-day mass of $\omega$~Cen.
This simple analysis is in good agreement with the simulations from \citet{bekki_freeman_03}, who found that a nucleated dwarf galaxy infalling from $26$~kpc onto a Milky Way-like galaxy can, through a combination of dynamical friction and tidal stripping of the host galaxy, result in a stripped nuclear cluster with an orbit similar to $\omega$~Cen.

The analysis also suggests $\omega$~Cen was deposited by the merger of its host galaxy at an apocentre no further than $\approx 9$-$11$~kpc (for a mass 1-3 times its present mass).
This is consistent with the energy `floor' found for G-E/S debris at $E \approx -1.8 \times 10^5 \kmsSq$ \citep{Horta_et_al_20}, very similar to the orbital energy of $\omega$~Cen of $E \approx -1.85 \times 10^5 \kmsSq$ \citep[assuming the potential from][]{McMillan_17}, suggestive of a causal connection.
The remaining uncertainty is why the orbit of $\omega$~Cen is less eccentric \citep[$e \approx 0.68$,][]{Baumgardt_et_al_19} than the bulk of G-E/S stars \citep[$e \approx 0.85$][]{Mackereth_et_al_19}.
However, we note that eccentricity may change during the course of a merger, such that merger debris at smaller apocentres becomes more/less eccentric than that at higher apocentres \citep[e.g. fig. 4 in][]{Pfeffer_et_al_20}.

Other tentative evidence comes from the NSC-to-galaxy mass ratios of nearby galaxies. A mass of $M_\mathrm{NSC} = 3.5 \times 10^6$~\msun\ suggests a host galaxy stellar mass of $\approx 10^9$~\msun\ \citep[][though with large scatter, and assuming the NSC-to-galaxy mass relation holds at $z>0$]{sanchez_janssen19}. Thus the more massive of the two suggested progenitors \citep[G-E/S:][]{Myeong_et_al_19, Matsuno_et_al_19, Kruijssen_et_al_20, Forbes_20} appears the most likely candidate.

Therefore, we assign $\omega$~Cen to be the NSC of G-E/S.

\subsubsection{M54}

Numerous works have provided kinematic and spatial evidence linking M54 with the Sagittarius Dwarf galaxy, including \citet{Ibata_et_al_94, Sarajedini_and_Layden_95} and most recently \citetalias{Massari_et_al_19}. M54 lies in the densest region of Sagittarius and has a distance and radial velocity consistent with the dwarf \citep{Da_Costa_and_Armandroff_95, Ibata_et_al_95, Ibata_et_al_97, Sarajedini_and_Layden_95, Layden_and_Sarajedini_00}.

\subsubsection{Terzan~5}

\citet{massari_et_al_15} were the first to measure a proper motion for Terzan~5, which they combined with the existing radial velocity measurements.  Integrating the resulting orbit of the cluster led the authors to conclude that there was no evidence for an ex-situ origin.  This was later confirmed by \citetalias{Massari_et_al_19} who found that Terzan~5 was kinematically associated with `main-bulge' component of the galaxy.  Hence, despite its characteristics of an age and metallicity spread, this appears to be an in-situ cluster.  We will discuss it in more detail in \S~\ref{sec:analysis_terzan5}.

\subsubsection{NGC~6273 (M19)}

\citetalias{Massari_et_al_19} unambiguously assign NGC~6273 to the `low energy' group of GCs.  This group has subsequently been associated with the Kraken merger event \citep{Kruijssen_et_al_19b,Kruijssen_et_al_20,Pfeffer_et_al_20,Forbes_20}.  Recently, Horta et al. (in preparation) have reported the discovery of the stellar component of Kraken, based on a combination of abundances and kinematics, lending further support to the existence of this relatively massive accreted galaxy.

As there are no other NSC candidates for this group, we assign NGC~6273 to be the NSC for Kraken.

\subsubsection{NGC~7089 (M2)}

While the reported spread in [Fe/H] at low metallicity within NGC~7089 appears to be related to an enhanced C+N+O population within the clusters (i.e., not a true [Fe/H] spread), there existence of a small population of significantly enhanced [Fe/H] is present within the cluster.

\citetalias{Massari_et_al_19} assign NGC~7089 to Gaia-Enceladus/Sausage based on its orbital properties.  As we have argued, $\omega$~Cen is likely to be the NSC of this accreted galaxy, hence it is unlikely that NGC~7089 is an NSC (see also the discussion in \S~\ref{sec:other_clusters}).

\subsubsection{NGC 6934}

Though it appears to have been accreted, \citetalias{Massari_et_al_19} do not associate NGC 6934 with a progenitor galaxy (allocating it to the `high energy' group of presumably unrelated GCs).
Its age and metallicity \citep[11.5-12~Gyr, $\FeH = -1.55$,][]{Dotter_et_al_10, VandenBerg_et_al_13} places it within the `accreted branch' of GCs, where a number of galaxy accretion events overlap in their age-metallicity relations \citep{Massari_et_al_19, Kruijssen_et_al_20}.
NGC 6934 has an angular momentum ($L_z$) and energy close to that of GCs associated with the \citet[hereafter H99]{helmi99} streams \citepalias{Massari_et_al_19}.
However its very high eccentricity \citep[$e \approx 0.9$][]{Baumgardt_et_al_19} more closely matches GCs associated with G-E/S.
If NGC 6934 was the NSC of the H99 streams progenitor, the question would remain of how it retained a higher energy than other H99 stream GCs (which would be unexpected for an NSC).
Alternatively, NGC 6934 could be another case of GC mass transfer (like we propose for M2, \S~\ref{sec:M2}) within its progenitor galaxy.

\subsection{Other Clusters}
\label{sec:other_clusters}

As discussed in \S~\ref{sec:constraints_iron}, there have been claims in the literature that other clusters may also be accreted NSCs.  After reviewing the evidence for each of these cases, we determined that most were not likely to be NSCs.  Here we look at the kinematic constraints from this group of clusters.  

Below we list these clusters as well as the kinematic group they are associated with, taken from \citetalias{Massari_et_al_19}.

\begin{itemize}
\item NGC~1851 - G-E/S
\item NGC~2419 - Sagittarius Dwarf
\item NGC~5824 - Sagittarius Dwarf
\item NGC~6656 (M22) - Main-Disk (non-accreted)
\item NGC~5286 - G-E/S
\item NGC~6864 - G-E/S
\item NGC~3201 - Sequoia or G-E/S (ambiguous)
\item NGC~6229 - G-E/S
\item NGC~6388 - Main-Bulge (non-accreted)
\end{itemize}

As noted by \citet{milone20}, many of these clusters (mainly Type II) are associated with G-E/S.
However, we do not expect G-E/S (or any other accreted system) to contribute more than one NSC to the Galaxy.
\citet{Pfeffer_et_al_14} found that the number of accreted NSCs a galaxy is expected to host correlates with the galaxy's halo mass. 
G-E/S had an approximate stellar mass of $\approx 10^{8.5}$~\msun\ \citep{Kruijssen_et_al_20, Mackereth_and_Bovy_20}, which corresponds to a halo mass of $\approx 10^{11}$~\msun\ \citep{Moster_et_al_18, Behroozi_et_al_19}.
We therefore expect G-E/S to contribute (on average) an additional $\approx 0.15$ accreted NSCs with masses $>10^5$~\msun\ \citep[equation 1]{Pfeffer_et_al_14}; i.e. dwarf galaxies will typically only contribute their own central NSC. 
This is consistent with the expectation that mergers are largely irrelevant for the formation and evolution of dwarf galaxies \citep[e.g.][]{Fitts_et_al_18, Davison_et_al_20, Martin_et_al_20}.
Therefore, under this interpretation, the other GCs associated with G-E/S are not likely to be NSCs.

Additionally, NGC~5824 and NGC~2419 belong unambiguously to the Sagittarius Dwarf, for which an NSC has already been identified (M54).  Hence we can rule these clusters out as NSC candidates, following the same reasoning above.

According to \citetalias{Massari_et_al_19}, M22 and NGC~6388 are kinematically associated with the Main-Disk and Main-Bulge, respectively, which would rule them out as accreted NSCs if they do not have an extragalactic origin.
In the case of NGC~6388, some claims have been made in favor of an extragalactic nature \citep{horta2020}. However, as NGC~6388 has a metallicity $\FeH \approx -0.5$ \citep{carretta_et_al_2007} and follows the in-situ branch of the age-metallicity relation \citep{Marin-Franch_et_al_09}, as well shows element abundances \citep[e.g. Al,][]{carretta_et_al_2007} higher than those typical of accreted objects \citep{das2020}, we consider the in-situ classification made by \citetalias{Massari_et_al_19} as the most likely.
M22 is a borderline case. It has a prograde orbit consistent with the disk \citepalias{Massari_et_al_19}, but is on a moderately eccentric orbit \citep[$e \approx 0.5$,][]{Baumgardt_et_al_19}.
It is also old and metal-poor \citep[$\FeH = -1.5$, age $12.7$~Gyr,][]{Forbes_and_Bridges_10}, thus in a region of age-metallicity space where the in-situ and accreted branches intersect.
Therefore it could either have been accreted, or an in-situ GC which had its orbit disturbed by galaxy mergers \citep[e.g.][]{Pfeffer_et_al_20}.
Given the lack of corresponding merger debris or other clearly accreted GCs on similar orbits to M22 \citepalias[e.g.][]{Massari_et_al_19}, we currently favour the latter scenario (i.e. an in-situ origin).

\section{Analysis - Pulling the Constraints Together}
\label{sec:analysis}

\subsection{Accreted Nuclear Star Clusters and Their Progenitor Galaxies}

By combining independent constraints obtained through stellar abundances and cluster kinematics we have found three clear cases of NSCs within the Milky Way GC population that we have been able to associate with known satellite accretion events.  The NSCs are NGC~5139 ($\omega$~Cen), NGC~6273 (M19) and NGC~6715 (M54), which were brought in by G-E/S, Kraken, and the Sagittarius Dwarf galaxy, respectively.

We found that the MW GC NGC~6934 displays some evidence for being an NSC, although further study is required to confirm its nature.  This cluster is potentially kinematically associated with the H99 streams.  Hence, it is possible that progenitor galaxy of these streams may have brought in NGC~6934 as an NSC.

We do not find any NSC candidate to be associated with the Sequoia accretion event.  The Sequoia, G-E/S, Sagittarius Dwarf and the H99 streams have estimated stellar masses (at the time of accretion) of $\sim1$-$3\times10^{8}$~\msun\ \citep[e.g.,][]{Kruijssen_et_al_20}. \citet{Neumayer20} have compiled a list of NSCs in the local Universe, along with the properties of their host galaxies.  They found that at these host galaxy stellar masses, $75-80$ per cent of galaxies host NSCs (though this may be a lower limit given the observational challenges in identifying NSCs). If we assume little evolution in the statistics (as the \citealt{Neumayer20} sample is collected at $z=0$ while the many of the MW satellites were accreted at $z>1$), we would therefore expect $\sim1$ of the MW accreted satellites not to host an NSC.  Hence, it may not be surprising that the Sequoia did not host an NSC.
The number of NSCs accreted by the MW is also consistent with the numbers expected from the assembly of MW-mass haloes \citep{Pfeffer_et_al_14, Kruijssen_et_al_19b}.

These results are summarised in Table~\ref{tab:nscs}.

\begin{table}
\caption{The candidate nuclear clusters and their most likely host galaxy} 
\label{tab:nscs}      
\centering  			
\begin{tabular}{c|c}
Accretion Event & NSC  \\ \hline
Kraken & NGC~6273 (M19) \\
\textit{Gaia}-Enceladus/Sausage & NGC~5139 ($\omega$~Cen) \\
Sequoia & none \\
Sagittarius & NGC~6715 (M54) \\
H99 streams(?) & NGC~6934 \\
 \hline 
\end{tabular}
\end{table}

\subsection{The Case of Terzan 5}
\label{sec:analysis_terzan5}

As discussed above, Terzan~5 hosts a complex stellar population with (at least) two sub-populations separated by $\sim0.5$~dex.  Additionally, \citet{ferraro_et_al_16} find that these two populations are separated in age, with the dominant ($60$ per cent, sub-solar metallicity) one having an age of $12.5$~Gyr and the minority component ($40$ per cent, super-solar metallicity) having an age of $4.5$~Gyr.

However, the kinematics of the cluster show that it is a member of the Bulge \citepalias{Massari_et_al_19} with an apocentre $\approx 2.8$~kpc \citep{Baumgardt_et_al_19}, hence does not have an extragalactic origin.

We note that the age-metallicity relation of the main components of Terzan~5 follow the trend expected for Galactic enrichment \citep[e.g.][]{Snaith_et_al_15, Kruijssen_et_al_19b}. This age-metallicity relation is also evidence against a galaxy accretion/NSC origin, since it would require a host galaxy of similar mass to the MW \citep[the MW's nuclear cluster is similarly metal rich,][]{Feldmeier-Krause_et_al_17} and a major merger $<4.5$~Gyr ago (i.e. more recent than the age of the youngest population) for which there is no evidence in the MW \citep[e.g.][]{Wyse_01, Hammer_et_al_07, Stewart_et_al_08}.

\citet{origlia_et_al_13} found evidence for a small sub-population within the cluster with $\mathrm{[Fe/H]} = -0.79$ which is also $\alpha$-enhanced.  Due to their small contribution to the total cluster mass, these stars may simply be accreted stars from the surrounding.

\citet{mckenzie_bekki_18} suggest that Terzan~5 may be the result of a bulge GC interacting with a GMC, accreting that gas and forming a 2nd generation in-situ.  This is possible, although if GC-GMC interactions were common, as suggested by the authors, then it is difficult to understand why more GCs in the Galactic central regions do not show such age/metallicity spreads.

Mergers of GCs appear to be rare in major galaxies, given the high relative velocities of the individual clusters. However, Terzan~5 has a disk-like prograde orbit \citep{massari_et_al_15, Baumgardt_et_al_19}, and merger rates are likely to be enhanced inside disks. In their simulation of a GC population inside a Milky Way-like galaxy, \citet{khoperskov_et_al_18} observed two major mergers of GCs inside the Galactic disk within 1.5~Gyr. Hence, it is conceivable that Terzan~5 represents a rare case of a cluster-cluster merger. Detailed studies of the internal structure and stellar kinematics of Terzan~5 may reveal further clues on the origin of this peculiar cluster. \citet{gavagnin_et_al_2016} predict that in the case of a merger, the structure of the final cluster will depend sensitively on the properties of the merging entities.

We note that a number of young massive clusters appear to have formed at the end of the Milky Way's stellar bar where it intersects with the Scutum-Crux Arm (e.g. \citealt{Davies_et_al_07, Alexander_et_al_09}; see also figure 3 in \citealt{Portegies_Zwart_et_al_10}).
This is consistent with the high molecular gas densities and star formation rates often observed at the ends of bars in other nearby galaxies \citep[e.g.][]{Downes_et_al_96, Sheth_et_al_00, Sheth_et_al_02}.
These clusters in the Milky Way have galactocentric distances of $\sim$3-5 kpc, close to the apocentre distance of Terzan~5 \citep{Baumgardt_et_al_19}.
Thus, if Terzan~5 happened to merge with a GMC or young massive cluster which formed at the end of the stellar bar around 4 Gyr ago (or less), this would explain why it is such an outlier in the Galactic GC population.

We conclude that the origin of the complex stellar populations within Terzan~5 is still unknown, although it appears that we can confidently rule out the possibility of it being an NSC.

\subsection{Intermediate-mass black holes}
\label{sec:imbh}

It is a long-standing question if intermediate-mass black holes (IMBHs, with masses $\sim10^3-10^5$~\msun) reside in some GCs \citep[see the recent review by][]{greene_et_al_2019}. In light of the ubiquity of supermassive black holes (SMHs) in the centres of massive galaxies and the well-established scaling relations between SMBH masses and galaxy properties \citep[e.g.][]{mcconnell_ma_2013}, the former NSCs of accreted galaxies appear as prime candidates to search for IMBHs. Further credibility for such a scenario comes from galaxies hosting both NSCs and SMBHs \citep[see the review by][and references therein]{Neumayer20} and the observation of SMBHs in massive ultra-compact dwarf galaxies (UCDs) observed around other galaxies \citep{seth_et_al_2014, Ahn_et_al_17, Afanasiev_et_al_18}, which are believed to be the remnants of accreted satellites \citep{Bekki_et_al_03, Drinkwater_et_al_03, Pfeffer_et_al_16}. However, evidence is still lacking for massive black holes in low-mass UCDs, which could be considered as extragalactic counterparts to $\omega$~Cen or M54 \citep{voggel_et_al_2018}. Confirming or refuting the presence of IMBHs in the former NSCs of the Galaxy is therefore crucial in order to understand if a lower mass limit for the formation of massive black holes in galactic nuclei exists.

For both, M54 and $\omega$~Cen, the presence of an IMBH has been suggested based on kinematic measurements\citep[e.g.][]{ibata_et_al_2009,noyola_et_al_2008}. However, in particular the case of $\omega$~Cen is still heavily debated \citep[see][and references therein]{baumgardt_et_al_2019b}. In both cases, further kinematic studies based on high-resolution data, such as HST astrometry or adaptive-optics assisted integral-field spectroscopy, will be required to answer the question if any of the clusters harbours an IMBH.

Alternatively, deep radio or X-ray observations can be used to search for signs of accretion of the intra-cluster medium onto an IMBH. To date, no such signals have been detected within the Galaxy \citep[e.g.][]{haggard_et_al2013,tremou_et_al_2018}, suggesting that IMBHs with masses $\gtrsim1\,000$~\msun are rare in globular clusters.

Our study suggests NGC~6273 and NGC~6934 as additional possible IMBH hosts. The central kinematics of both cluster have not been studied with high-resolution data so far. Another way of constraining the presence of massive black holes is by searching for signatures of gas accretion in deep radio data. As neither NGC~6273 nor NGC~6934 were included in previous radio surveys of Galactic GCs \citep[e.g.][]{tremou_et_al_2018}, they appear as promising future targets in the hunt for IMBHs. 

Besides M54 and $\omega$~Cen, a number of Galactic GCs that we did not identify as likely former NSCs were suggested to host IMBHs \citep[see compilations by][]{baumgardt_2017,greene_et_al_2019}. However, all reported detections were not confirmed in follow-up studies and are therefore still controversial. We note that the detection of an IMBH in a GC does not necessarily imply that the cluster formed as an NSC, given that some of the mechanisms proposed to form massive black holes do not require the cluster to sit in the centre of a galaxy \citep[e.g.][]{gieles_et_al_2018}.

\section{Summary}

The goal of this study has been to find the Nuclear Star Clusters (NSCs) that have been accreted by the Milky Way, and associate them with their progenitor galaxy.  We began with a sample of 15 GCs which have been claimed in the past as possible accreted NSCs.  We have applied two independent constraints to assess the possibility of each cluster being an NSC, namely the internal abundance spreads (specifically [Fe/H]) and the orbital properties of the cluster.

From an analysis of the abundance spreads we found six GCs with clear evidence of an internal [Fe/H] spread.  They are $\omega$~Cen, M54, NGC~6273, Terzan~5, NGC~7089 and NGC~6934. 
Given the lack of detailed spectroscopic studies for many Milky Way GCs (particularly those nearest the Galactic centre), it is possible more candidates will be discovered in the future.  While NSCs are all expected to host Fe-spreads within their stellar population, a handful of known MW GCs, that are not strong NSC candidates, also host such spreads.  The origin of these Fe-spreads is currently unknown and is a rich avenue for future studies.

By looking at the orbital properties (and their origin, either in-situ or ex-situ) we found four NSC candidates that can be associated with an galactic accretion event, hence are likely to be genuine NSCs.  These are listed in Table~\ref{tab:nscs}.  Of the five identified main accretion events (Gaia-Enceladus/Sausage, Sequoia, Kraken, Helmi-streams, Sagittarius dwarf galaxy), we find an associated NSC for all except the Sequoia event (and possibly also the Helmi-streams).  In the inferred mass range of the accreted satellites ($\sim10^{8}$~\msun), NSCs are found in $\sim80$ per cent of galaxies in the local Universe, in good agreement with the statistics implied by our results.
These four former/current NSCs ($\omega$~Cen, M54, NGC~6273, and NGC~6934) are the best candidates for searches of IMBHs within star clusters of the MW.

We found that there were two GCs that host significant internal iron spreads that are unlikely to be accreted NSCs, namely Terzan~5 and NGC~7089 (M2).  The origin of these iron spreads is currently unknown.  Perhaps these clusters represent rare events of cluster-cluster mergers, collisions with molecular clouds or the accretion of stars from one cluster to another due to a close passage.

\section*{Acknowledgments}

We thank GyuChul Myeong and Chris Usher for helpful discussions and suggestions.
We thank the referee, Gary Da Costa, for his careful and constructive report that helped improve the paper.
JP, NB, SS and SK gratefully acknowledge financial support from the European Research Council (ERC-CoG-646928, Multi-Pop). NB gratefully acknowledges financial support from the Royal Society (University Research Fellowships). 
CL gratefully acknowledges financial support from the European Research Council (ERC-2018-ADG STAREX).

\section*{Data Availability}

No new data were generated or analysed in support of this research.



\bibliographystyle{mnras}
\bibliography{bibliography}


\bsp	
\label{lastpage}
\end{document}